\documentclass[conference]{IEEEtran}
\usepackage{cite}
\usepackage{amsmath,amssymb,amsfonts}
\usepackage{algorithmic}
\usepackage{graphicx}
\usepackage{textcomp}
\usepackage{xcolor}
\usepackage{enumitem}
\usepackage{url}
\usepackage{wrapfig}

\graphicspath{./images/}

\title{ 
        	Cloud Native Software Engineering 
      }

\author{
			Brian S. Mitchell\\
			Department of Computer Science\\
           College of Computing and Informatics\\
			Drexel University, Philadelphia, PA, USA\\
			bmitchell@drexel.edu
} 

\date{}

\begin{document}

\maketitle

\begin{abstract}

Cloud compute adoption has been growing since its inception in the early 2000's with estimates that the size of this market in terms of worldwide spend will increase from \$700 billion in 2021 to \$1.3 trillion in 2025\cite{IDCReport}. While there is a significant research activity in many areas of cloud computing technologies, we see little attention being paid to advancing software engineering practices needed to support the current and next generation of cloud native applications.  By cloud native, we mean software that is designed and built specifically for deployment to a modern cloud platform. This paper frames the landscape of Cloud Native Software Engineering from a practitioners standpoint, and identifies several software engineering research opportunities that should be investigated. We cover specific engineering challenges associated with  software architectures commonly used in cloud applications along with incremental challenges that are expected with emerging IoT/Edge computing use cases.

\end{abstract}

\section{Introduction and Context}
\label{Intro}
Delivering managed computing services on hosted infrastructure started in the late 1990's with the introduction of the Software-as-a-Service (SaaS) model. One of the early pioneers of this model was Salesforce.com\cite{SalesforceHistory}, which launched in 1999.  Unlike other companies that licensed software  deployed on customer-owned equipment, SaaS companies provide a pay-as-you-go  subscription model. In this model, they manage all of the software and compute infrastructure, you pay a monthly charge that entitles access to the solution from any device at any time.     

While SaaS solutions marked the start of shifting software license spend to usage-based spend, public cloud computing as we know it today can be attributed to the launch of AWS (Amazon) \cite{AWSLaunch} in early 2006, with Azure (Microsoft)\cite{AzureLaunch} and GCP (Google)\cite{GCPLaunch} following in 2008. The primary early adopters of cloud computing were technology companies that innovated patterns, practices, and open sourced tools and frameworks that have become best practices for running resilient and scalable business services in the public cloud.  Over the past 10 years cloud computing has been growing in organizations of all sizes across many different industries. 

Larry Wall, the creator of Perl, once stated -- \textit{There is a saying in the software design industry: ``Good. Fast. Cheap. Pick two."}.  Software engineering involves making difficult decisions based on informed tradeoffs.  For example, it would not be hard to argue that in order to move faster and build things cheaper, compromises on software features, software quality, and/or security would be required. Using the utility of the cloud, coupled with modern cloud computing tooling, one can now argue that you can build better software faster and cheaper.  It's not that Larry Wall's insights were incorrect, but we can now have the technologies and practices to redefine \textit{good} in terms of \textit{fast} using the cloud.  When computing components are deployed to the cloud, the simplest way (and thus the most popular way) to do this is via automation\cite{terraform, AWSCloudFormation, AzureLaunch, Pulumi}.  The \textit{automate everything} practice embraced by cloud computing not only allows deployments to be fast, but it also favors ephemeral computing components. These components by their nature are easier to test\cite{kim2016devops} and can be started, stopped, paused, or replaced at any time. 

This combination of capabilities enables software engineers to rapidly deploy software to a known state at any time. With these building blocks new well-tested features can be quickly and consistently rolled out to users in very small batches.  Goodness of the solution can now be validated via feedback from users, either directly, or via monitoring and instrumentation of their behavior.  These cloud enabled capabilities have the potential to advance software engineering practices in many ways, but transforming these practices across the entire community comes with many challenges. We think this represents a significant opportunity for the software engineering field given the likelihood that most industrial systems moving forward will be deployed on cloud runtimes\footnote{By cloud runtime we include public, private and hybrid cloud infrastructure}. Specifically:

\begin{enumerate}
	\item Helping Software Engineers manage the expanded cognitive load required to design, build, deploy and operate at scale  applications in the cloud. We will discuss this throughout the remaining sections of this paper. 
	
	\item Identify opportunities to accelerate and scale software engineering skillsets needed to deploy a broader suite of applications to the cloud. Many organizations will want to move beyond deploying externally facing web and mobile applications to the cloud using their top engineers. This will require developing new skills for the broader engineering organization as more of their core business moves to cloud computing.
	
	\item Investigate how software engineering and computer science education can expand to address the demands of industry to create new, and retool existing software engineers for the cloud.\footnote{We will talk about cloud certifications later, but they are targeted towards using the services of a cloud provider, not on the design and architecture of cloud native applications} Most cloud-proficient software engineers appear to build their skillsets on the job and with online resources versus in formalized academic programs. 
	
	\item Understand software engineering needs for new architectures enabled by the cloud.  For example, IoT and smart devices that run at the edge increase the complexity of software engineering given the  distributed nature of these platforms.  These challenges will be discussed in Section \ref{subsec:Edge}.  
	
	\item Address non-technical challenges that organizations face with adopting cloud-centric engineering best practices. Consider Google who published in 2021\cite{GoogleDevOps} that they ran over 700K experiments in production that resulted in over 4K search product changes. Netflix open sourced tools\cite{NetflixChaos} that they use for chaos testing to validate platform resiliency.  Comfort with strategies that involve testing and randomly breaking things in production are embedded in the DNA of technical companies, but are often met with caution in traditional organizations. 
\end{enumerate}

We will address a number of these opportunities in the subsequent sections of this paper.  The next section will introduce \textit{Cloud Native} from a software engineering vantage point. Throughout this paper, by cloud native, we are referring to systems designed specifically to favor managed cloud platform services (PaaS/Faas)\cite{albuquerque2017function}, and not systems that are \textit{lifted and shifted}\cite{CloudMigration2017} from an on premise virtual machine to a virtual machine that runs in the cloud (IaaS).

\section{What is Cloud Native Computing?}
\label{sec:WhatIsCNF}
Before we explore the software engineering landscape for the cloud, we need to address exactly what we mean by cloud native computing.  According to the Cloud Native Computing Foundation (CNCF)\cite{CNCFHome}  \textit{``Cloud native technologies empower organizations to build and run scalable applications in modern, dynamic environments such as public, private, and hybrid clouds."}.  Amazon's definition is \textit{``Cloud native technologies empower organizations to build and run scalable applications in modern, dynamic environments such as public, private, and hybrid clouds"}. Google offers the definition \textit{``Cloud native means adapting to the many new possibilities—but very different set of architectural constraints—offered by the cloud compared to traditional on-premises infrastructure."}.  The primary theme in these definitions centers around the role that cloud platforms play in enabling the creation of cloud native applications.  They also don't clearly define ``Cloud Native", which we consider any application that is specifically designed to be deployed on a cloud platform. 

We think a better definition of cloud native computing that focuses more on  software engineering is \textit{``Cloud native applications are well architected systems, that are ``container" packaged, and dynamically managed"}. Specifically:

\begin{figure*}[t]
	\includegraphics[width=\textwidth]{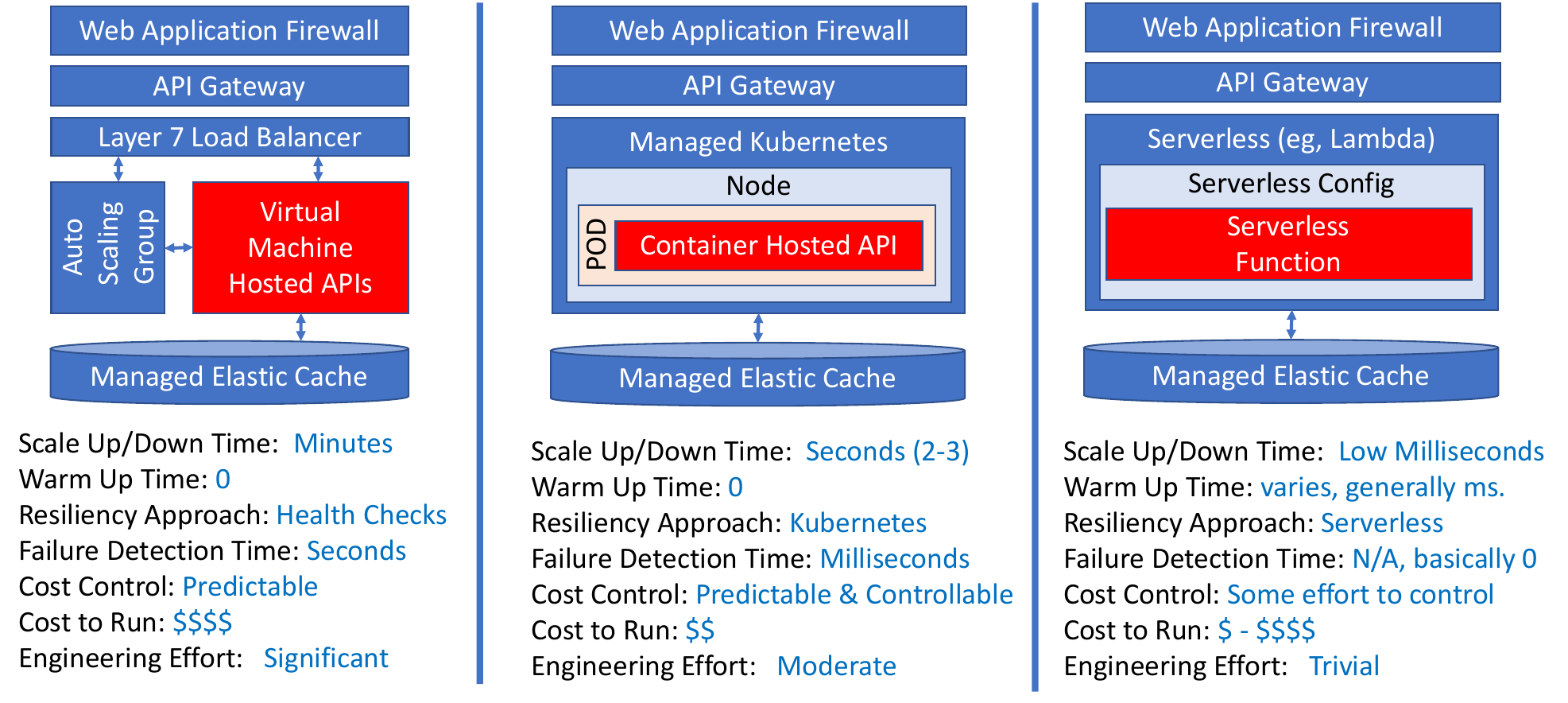}	
	\caption{Cloud Native Architectural Tradeoffs for a Hypothetical Suite of APIs}
	\label{fig:CloudQATradeoffs}
\end{figure*}

\textbf{Well Architected Systems} - By this we mean systems that adhere not only to established software engineering best practices but also embrace specific functional and non-functional capabilities offered by the cloud. For example, how the computing components such as services/APIs are identified, how they work with each other, how security requirements are met, and how the system is designed for resiliency and scale?

\textbf{Container Packaged} - The term \textit{container} is overloaded in the cloud computing terminology landscape.  In many places its equated to a standardized package\cite{OCIStandard} that is managed by Docker\cite{DockerContainer} technologies - aka ``a docker container".  We take a more generic view of container packaging. Specifically, we think container packaging is a mechanism to package and deploy code that is ephemeral, can operate across a variety of different hardware architectures (e.g., Intel, ARM, microcontrollers, etc), and at runtime is supervised.  Supervision includes full lifecycle management associated with version identification, startup, shutdown, health checks, security scanning, and monitoring.  Examples of container packaging and supervision include Docker, Docker Compose, Kubernetes, and serverless \cite{baldini2017serverless}. We also include in this category the emerging interest with using server-side web assembly\cite{haas2017bringing, bosshard2020use} as a way to package and deploy cloud native application services. 

\textbf{Dynamically Managed} - Consider the cloud as a large, highly distributed, special purpose operating system. Just like any operating system, there are a number of resources like storage, compute, network and security services that are needed by applications.  The job of an operating system is to dynamically manage and optimize the allocation of these resources to the realtime computing demand on the overall system.  When done well, every process being managed by the OS will perceive that it has access to the resources it needs, when it needs it.  In a similar context, a cloud service provider, via Application Programming Interfaces (APIs), provides and manages resources to cloud native applications dynamically. Classical operating systems manage physical resources on a single system, whereas cloud resources are virtualized and distributed, while also being resilient and scalable.  For example, block storage that supports virtual machine reads and writes are automatically replicated across servers in different special purpose data centers. Outside of initial configuration, the user does not worry about how durability is provided given its dynamically managed by the cloud service provider. Other examples include using auto scaling of virtual machines with health checks, or more advanced services like Kubernetes\cite{kubernetes} that can scale up or down dynamically based on demand. Function as a service (FaaS) solution's take this a step further by running code on demand when a certain event happens.  AWS even open sourced a micro VM called Firecracker\cite{AWSFirecracker} they created to support dynamically managing serverless workloads at scale.   

\begin{figure*}[t]
	\includegraphics[width=\textwidth]{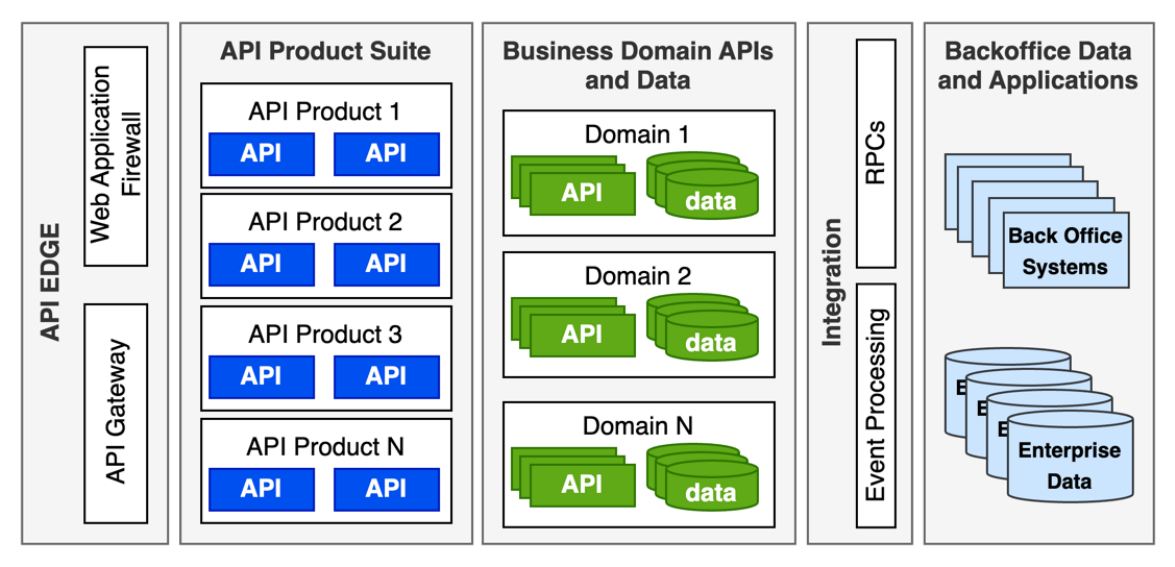}	
	\caption{Conceptual API-Based Architecture}
	\label{fig:APIArchitecture}
\end{figure*}
Now that we have provided a definition, we describe next a number of interesting software engineering problems that warrant investigation.

\textit{Managing cloud native technical assets}.  In 2011 Adam Wiggins authored a set of technical principals that enable software engineers to create, manage and release code in support of cloud native applications. These principals were branded ``The Twelve-Factor App"\cite{12factor}.  Over the years they were updated and revalidated\cite{hoffman2016beyond, 12factorRevisited}, but consistently hold up as recommended engineering practices. One of the key challenges is that while none of these practices is overly complex, they require mastery and discipline. Also, existing standards enforced by enterprises might act as blockers to some or all of these practices.  For example, some organizations might not have comfort or necessary controls to ensure that all deploys, to all environments, are driven through a source code control system.

\textit{Identifying an appropriate cloud native software architecture}.  The technical principals associated with 12 factor apps is a good start; however, these only focus on how to manage cloud native technical assets.  Good cloud native solutions are generally architected as a suite of horizontally composable components. This model introduces several interesting challenges for cloud native software engineers that must be addressed. What are these components? How do identify them? What will guide how they should be built? We think some of the newer concepts around app meshes\cite{GartnerMASA} and data meshes\cite{datamesh} provide a good start for shaping the overall architecture.  As far as identifying the architecture components themselves, we think concepts from Domain Driven Design (DDD)\cite{evans2004domain, vernon2013implementing} can be refreshed to support this activity.

\textit{Upskilling Software Engineers on the use of Quality Attributes to make informed technology tradeoff decisions}. Most cloud providers offer many different technology choices to create cloud native components.  Applying discipline around quality attributes should guide making technology stack decisions. For example, Figure\ref{fig:CloudQATradeoffs} shows three different ways to deploy cloud native APIs along with some associated quality attributes.  It should be noted, that the values of these quality attributes will change for different APIs, the figure assumes these hypothetical APIs are very light weight, event triggered, and manage their state via an elastic cache.  Thus for the scenario shown in the figure, the VM option on the left does not make sense given the high cost, large scale up/down times, and complex engineering effort.  The container option in the middle and the serverless option on the right both seem like good options.  The ultimate decision will be driven by the desire to have a high degree of control over cost, and the nature of the workload. For example, if the expected traffic has massive near realtime spikes, a serverless solution might be preferred. If the workload is not event-driven, or has many runtime dependencies such as database connections, then a container based solution might be a better choice. As we move more towards computing at the edge, containers might not be possible since they depend on Linux kernel, so emerging lighter weight alternatives such as WebAssembly (WASM) with WASI\cite{WASI} might be a good choice.  Cloud native software engineers must be able to make these types of choices and resist over-standardizing based on personal or organizational preference and let software architecture practices using quality attributes guide cloud native platform choices. 

\textit{Organizing teams for cloud native success}.  In 1967, Mel Conway published a paper called ``How Do Committees Invent" - Fred Brooks cited this paper in the Mythical Man Month\cite{Brooks1975} calling it Conways law\cite{ConwaysLaw}. Conways law states \textit{``Any organization that designs a system (defined broadly) will produce a design whose structure is a copy of the organization's communication structure"}.  In the early days of Amazon, Jeff Bezos introduced the idea of the ``2 pizza team rule"\cite{TwoPizza} where a team size should be no larger than can be fed by two pizzas.  The basic ideas are rooted in concepts that productive teams should be small, and independent.  This aligns nicely with cloud native concepts in that one way to ensure that components are independent and interoperate only via their published interfaces, is that teams are also organized this way.  Over the years various organizational models and changes have been debated. \textit{Full-stack teams} bring together front-end, back-end, testing and infrastructure professionals to a common team where they have full responsibility for their technical assets. \textit{Shifting left} along with the emergence of \textit{DevOps} brings a testing and automation focus to teams that allows them to increase quality and productivity.  One problem is that while these concepts work well for driving individual team productivity, they are difficult to scale to larger organizations with multiple products.  Several companies have also published their attempts to scale their practices.  One popular model was published by Spotify\cite{SpotifyModel}, where ``Squads" represent full-stack teams, but they also introduce concepts like ``Tribes" for coordinating squads, and ``Guilds" to address cross-cutting technical concerns. Commercial models have also been created to drive organizational changes to support cloud native architectures.  One such example is the SAFe\cite{SAFeAgile} framework, which is interesting in that it favors prescriptive organization and process rigor over streamlining software engineering practices in order to increase scale. We think there are some interesting problems in this space given the lack of alignment on the best ways to organize teams to work on cloud native applications at scale. Specifically, just copying a model used in one organization and using it in another organization does not address many of the challenges associated with things like politics and culture.

\begin{figure}[b!]
	\includegraphics[width=\columnwidth]{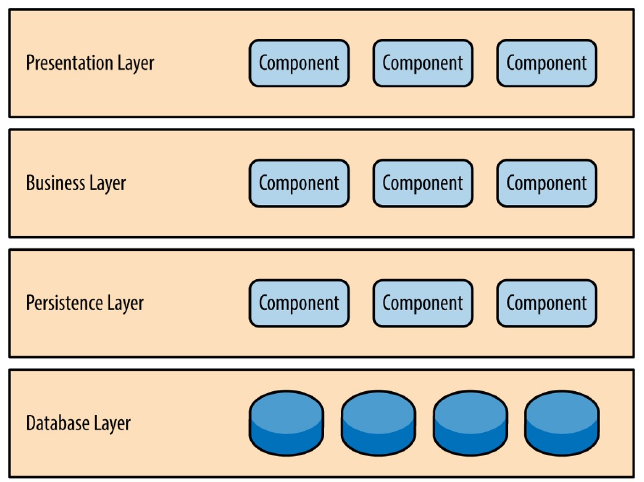}	
	\caption{Layered Software Architecture}
	\label{fig:LayeredArchitecture}
\end{figure}

\textit{Software engineering demands for API based technical products}.  Historically, most applications are built to solve a targeted user or business problem end-to-end.  In Mark Richards book entitled ``Software Architecture Patterns"\cite{richards2015software}, Chapter 1 is focused on the \textit{Layered Architecture Pattern}, which is shown in Figure \ref{fig:LayeredArchitecture}. Richards layered pattern expands on the 3-tier architecture pattern\cite{aarsten1996patterns} that calls for the isolation of the presentation, business logic, and database layers. An important attribute of these patterns is that the end-user interfaces only with the presentation layer, allowing the remaining layers to be hidden and secured against direct access. One of the foundational enablers of cloud computing is the plethora of first-class integration services provided by the platform. These capabilities open the door for new software architectures based on offering API-enabled services as a product. We propose a conceptual model for this type of architecture in Figure \ref{fig:APIArchitecture}, while it is layered, the layers have different responsibilities than the ones discussed by Richards. Several well known examples of API products are Google Maps (for location services), Twilio (for messaging), and Stripe (for payment). These all represent API enabled capabilities designed specifically for embedding into other applications. As this model expands in popularity software engineers will have to become familiar new architecture patterns for designing API-centric products.  For example, some of the best AI models are offered as a service by OpenAI\cite{OpenAI}, and the healthcare industry is being mandated to offer API services to support interoperability\cite{FHIRAPI} regulations.  Some of these companies are even taking the opportunities to use APIs as a competitive differentiator, one example is the developer portal provided by Cigna\cite{CignaDeveloper}. Prior to the mainstream adoption of the cloud native applications it was not possible to work with your bank, healthcare provider, auto insurer, and favorite retail stores with custom developed software.

\begin{figure*}[t!]
	\includegraphics[width=\textwidth]{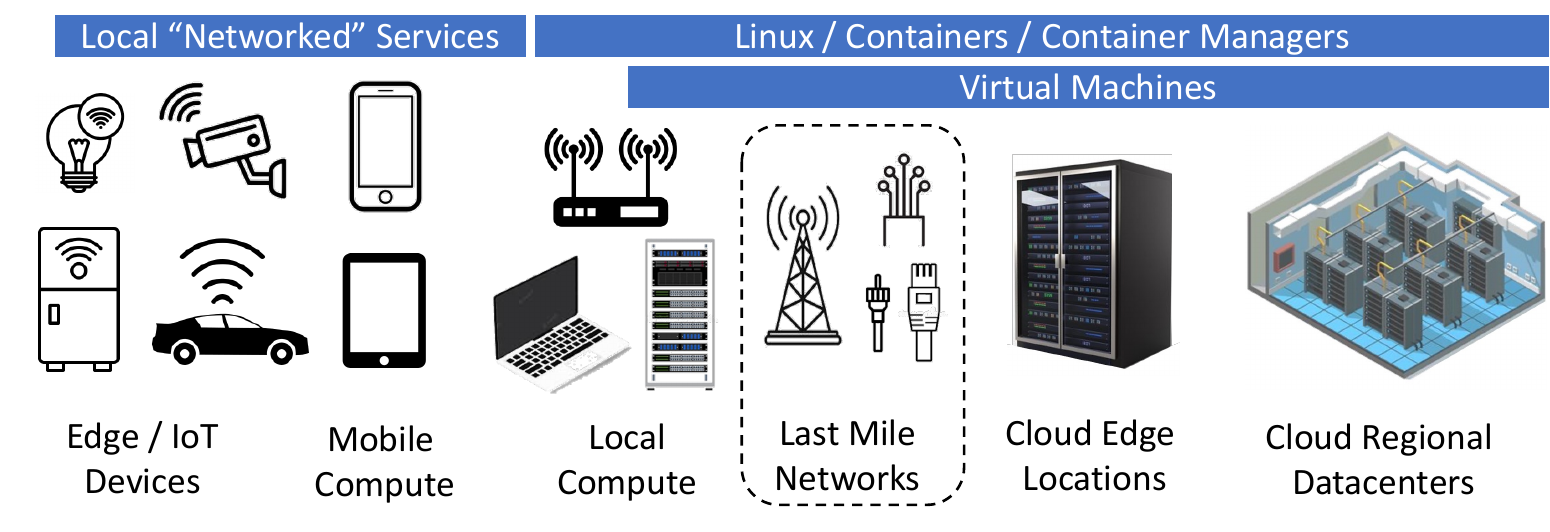}	
	\caption{The Modern Cloud}
	\label{fig:CloudTopo}
\end{figure*}

\section{The Cloud is Expanding to the Edge}
\label{sec:CloudArchitecture}

The underlying services that the major cloud providers offer to their customers continues to expand and evolve.  These advancements provide significant innovation capabilities to customers, but also put pressure on how to effectively engineer solutions that take advantage of these services cost effectively.  Figure \ref{fig:CloudTopo} shows the high level view of the computing services in the modern cloud.  While it was once easy to identify the boundary of where the cloud started and ended, it is no longer easy to define the edge of the cloud.  The following section will provide an overview of the evolution of the cloud, along with highlighting some of the challenges that have to be overcome by software engineers.

\subsection{The Basic Cloud Architecture}
\label{subsec:BasicCloudArchitecture}

At a high-level the basic cloud architectures deployed by major providers exhibit many similarities.  The foundational infrastructure building block of cloud compute is called an \textbf{Availability Zone} (AZ).  An AZ is a custom designed data center that hosts cloud infrastructure (compute, storage, and network) and runs cloud provider services on behalf of their customers.  A \textbf{Region} is a physical location where a collection of 2 or more AZs are located.  Each AZ within a region are connected together with a fully redundant high bandwidth low latency network.  The goal of a region is to have AZs close enough so that they can behave like a single cluster, but also separate them by enough distance to isolate them from issues associated with power failure, earthquakes, tornados, and so on.  AWS, as an example, uses the general guideline of 100km (60 miles)\cite{AWS-AZ} for placing AZs within a region. The global footprint of a cloud provider is defined by the number of regions and locations they have across the globe, along with the purpose-built underlying network they use to interconnect them together. 

Given the cost and complexity of deploying cloud regions around the globe, cloud providers expand their network reach through the use of edge locations.  Edge locations are useful for a couple of reasons. First, they serve as a point of presence to lower connection latency to the cloud, and second, they can run services at the edge which offers additional benefits.  One of the classical applications to run at the edge is a content delivery network (CDN). CDNs speed up web and mobile applications. For digital web and mobile applications the combination of Regions, AZs, and Edge Locations could be considered the boundary of the cloud. These are shown on the right side of Figure\ref{fig:CloudTopo}.  The major cloud providers continue to focus on expanding the number of services they support at the edge to drive even better performance and reduce network latency.     

From a software engineering perspective, the basic cloud architecture described above introduces additional cognitive load on software engineers:

\begin{itemize}
	\item  Planning application deployment starts with the design of a virtual data center (VDC). VDCs logically carve out storage, compute, network and security policies from the cloud provider for customer usage. Traditional software engineers are not trained to think of the start of the software design process begins with the need to design a virtual data center and all of the complexity that comes with it.  Historically, data centers are designed by specialty engineers and inherited ``as is" into the final software architecture. The data center topology is now a critical software engineering concern. 
	
	\item  Quality attributes such as privacy, resiliency, reliability and scalability are foundational concepts that software architects use to reason about systems.  These now move out of the conceptual realm and require a deeper understanding of technical constructs that now become part of the software design itself.  For example, deploying microservices across different subnets, where each subnet is in a different AZ within a region. Also, declarative definition of  the security policies that govern access to these microservices along with their entitlements to access other cloud resources becomes part of the software product itself. 
	
	\item  Although the cloud itself provides an infrastructure model to create resilient solutions that run at scale, its up to the software engineer to architect things properly to take advantage of these capabilities\footnote{Some patterns such as rehosting\cite{engelsrud2019moving} {\em a.k.a.} ``lift-and-shift" should not be considered cloud native patterns.}.  For example, it's still possible to deploy an application to a single virtual machine instance in the cloud, which without additional controls will not elastically scale, nor will it be resilient to failure.  Thus, to enable the creation of cloud services software engineers must have mastery of newer patterns for distributed applications ( {\em e.g.}, especially asynchronous event-based architectures).
	
	\item  While cybersecurity has always been an important consideration of software engineers, the cloud materially expands these responsibilities. Everything in the cloud is secured by policy, but as mentioned earlier, software engineers now need to deal with security requirements across the entire OSI model\cite{OSIModel} stack in addition to some unique cloud requirements. Generally software engineers are comfortable with security at Layer 7 (the Application Layer) of the OSI model, using techniques such as OAuth 2\cite{oAuthStandard} to secure different types of digital assets. These responsibilities now expand to authoring and deploying policies to govern network access across subnets, and for attaching policies necessary to use managed services.  In addition, software engineers must deploy and ensure proper configuration of virtualized security appliances such as web application firewalls (WAFs) and traditional firewalls that are now virtualized and software-defined. As attacks get more sophisticated, software engineers must also make decisions around introducing additional security capabilities into their solutions such as bot-detection, and defenses against credential-stuffing via MFA and supply chain attacks.  The complexity of properly configuring, keeping track of, and managing cloud resources is also a new cloud-specific concern for software engineers.  These problems themselves are also being addressed by software which needs to be deployed, configured and managed; for example Cloud Custodian\cite{CloudCustodian} that was open sourced by Capitol One and donated to the CNCF. 
	
	\item  One clear benefit of deploying to the cloud is that the easiest path to do so requires everything to be automated. While software engineers are comfortable with automation associated with software tasks like testing, they are not accustomed to automating infrastructure deployment. This becomes even more challenging given many of the existing infrastructure automation tools were designed for non-programmers, relying on verbose, complex and error-prone configuration formats like YAML and JSON. Some progress in this space has been accomplished via DSLs like Terraform\cite{terraform} and tools that use real programming languages like Pulumi\cite{Pulumi}. 
	
	\item  Another foundational software engineering cloud concern that design decisions have material influence on operational runtime costs. This is often referred to as \textit{finops}, short for financial operations. At its core, the cloud transforms compute, network, storage, and security into a pay-as-you-go utility.  Software engineers generally don't factor in things like programming language selection, database platforms, processor hardware architecture, frameworks, fully-managed services and so on into their design from the perspective of cost and carbon footprint impact.  We will explore this topic more in Section \ref{sec:Polyglot}.
	
\end{itemize}

\subsection{The Emergence of Edge Computing}
\label{subsec:Edge}

With the rapid growth of devices that are connected to the internet, we are now entering the era of edge computing\cite{edgecomputing}. Edge computing is  different architecturally from traditional cloud computing.  Consider a cloud-enabled web or mobile application.  The architecture of these applications is often based on calling cloud-hosted APIs and then using the data returned from these to power the user experience.  This architecture will not scale or meet the needs of all of the smart devices that connect to the internet.  As its name implies, edge computing moves more computing services to the edge, with requirements not found in web or mobile applications:  Specifically:

\begin{itemize}
	\item They must be able to work autonomously. The cloud would not scale to support all device events, local processing is used to filter important events from less important events.
	
	\item They must be able to work fully disconnected, or with unreliable network connectivity.  
	
	\item They must be able to perform compute locally, either independently, or in local clusters. 
\end{itemize}

\begin{figure}
	\includegraphics[width=\columnwidth]{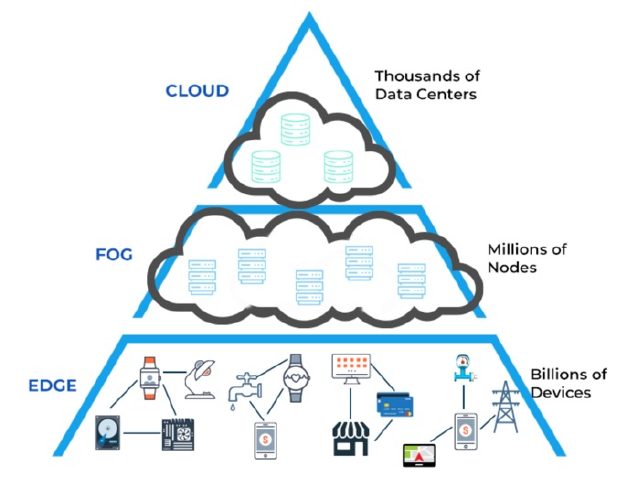}	
	\caption{Edge Compute Architecture}
	\label{fig:EdgeArchitecture}
\end{figure}

These added capabilities essentially extend the edge of the cloud all the way back to the client devices themselves as shown in Figure \ref{fig:CloudTopo}. The overall architecture of the modern cloud that extends to the edge is shown in Figure \ref{fig:EdgeArchitecture}\footnote{Figure copied from \url{https://www.spiceworks.com/tech/cloud/articles/edge-vs-fog-computing/}}.

According to research by IoT Analytics, there were 14.4 million smart devices connected to the internet in 2022, expected to rise to almost 30 million by 2025\cite{IoTMarket}.  This many deployed devices could not be supported if they required connectivity to the large cloud data centers discussed earlier. Processing will need to move to the devices themselves supported by a new layer of cloud compute that is closer to the devices.  This new layer of compute is often referred to as \textit{fog computing}.  The term comes from a play on the word cloud, given clouds are high up in the sky and fog is closer to the ground.   

As cloud providers expand their footprint across the globe, creating new regions with multiple availability zones represents a major strategic decision because of the time, expense and other factors that go into rolling out multiple large data centers. Creating new regions is required to increase compute capacity as global cloud adoption expands, and to meet specific compliance requirements associated with conducting business in the cloud. For example, many countries are adopting data residency laws, which place controls over where data is stored at rest. The trend of cloud providers searching for strategic locations for new Regions, or to expand the number of AZs within a region will continue as cloud demand increases. Given the massive investments required, it is likely that the number of major cloud providers will continue to be small.  A November 2022 TechTarget report\cite{CloudMarketShare} highlights that the big 3 providers -- AWS, Microsoft, and Google -- account for 62\% of the overall cloud market. 

While its unlikely that there will be significant disruption in the major cloud providers, the fog layer is likely to be federated across many different players.  This layer needs to be deployed close to the edge devices themselves, and will likely be addressed by existing last mile internet service providers (ISPs), and by telecommunication companies offering 5G services who already have deployed infrastructure to meet these needs.

\section{The Edge and Expansion of Software Engineering Concerns}
\label{sec:Polyglot}
The evolution of cloud computing over the past decade has increased the decision landscape for software engineers. This section will highlight some of the new concerns that software engineers must address in cloud and edge computing design.

\subsection{Processor Hardware Architecture Diversity}
Ten years ago we did not have cloud providers creating custom processors for compute, special purpose AI applications, nor did we have all of the microcontrollers running at the edge of the Internet. Familiarity with making informed hardware architecture choices now becomes an important concern of software engineers. Some examples include:

\begin{itemize}
	\item On May 23, 2023, AWS announced the third generation of their custom ARM-based microprocessor called Graviton 3.  AWS claims that workloads running on Graviton 3 are 50\% faster than Intel/AMD processors, consume 60\% less power, and are 20\% cheaper. From a software engineering perspective these benefits seem like a no brainer to take advantage of until you start to factor in other requirements such as being able to maintain ARM-based builds of your software, including all dependencies which may not be available or optimized for ARM. Additionally, organizations may impose other requirements such as running certain security products on VMs, these also must be available and certified. 
		
	\item Since the realization that GPUs can improve the performance of training AI models, cloud providers have innovated further with custom AI microprocessors.  In 2016 Google introduced the Tensor Processing Unit (TPU) to accelerate training deep learning models, and in 2018 AWS created the Inferentia chip to accelerate inference. AWS also entered the training space to compete with Google's TPU with the Trainium chip in 2020.  With all of these new AI hardware choices, software engineers must be savvy with aligning hardware choices with software training and inference library requirements. For example, AWS announced a SDK for Trainium called Neuron to enable engineers to use popular AI frameworks such as Tensorflow and PyTorch.  
	
	\item As we move to the edge, software engineers now more routinely have to create solutions for microcontollers, and other devices that have additional constraints.  These devices might be battery powered, compute constrained, difficult to access or update, and/or have unreliable network connectivity. Programming frameworks and tools routinely available on modern servers might not be available or viable for these devices.  Consider the popular trend of deploying code in containers.  To a large extent, containers make assumptions that there is an underlying linux kernel, which might not be possible in these purpose-built devices. Instead of falling back to creating alternative versions of their software in lower level systems programming languages like C/C++, software engineers must become familiar with emerging solutions in this space.  Consider TinyGo\cite{TinyGo}, which is an alternative Go compiler specifically created to bring the Go ecosystem, which is popular in the cloud, to microcontrollers.  Another example, is WasmEdge\cite{WASMEdge}, which brings the power of Web Assembly(WASM) to the server and to edge devices.  WasmEdge can run embedded WASM code created by modern compilers that are popular for creating cloud native applications such as Rust, Go, and Javascript.       
	
	\item Managing tool chains for multiple hardware architectures.  The Java programming language introduced the concept of \textit{``Write Once, Run Anywhere"}. It accomplished this by creating Java Virtual Machines (JVMs) for different hardware platforms, and running compiled bytecode consistently across these platforms.  While this works well, the Java ecosystem has some challenges in the broader cloud native space.  Specifically, to use Java all dependencies must be Java-based, and although the JVM itself is an impressive, its size and compute requirements might be challenging to support on edge devices. Newer programming languages like Rust and Go have been adopting an open cross-compiler philosophy so that any compiler on any platform can create binaries for any other platform.  Containers are another popular cloud native technology.  With the need to support diverse processor architectures container packaging becomes more complex, and containers might not be practical on the edge given they assume the presence of a linux kernel.  Docker recently released a technical preview to support web assembly that may help address this issue\cite{DockerWASM}.
\end{itemize}
 
\subsection{Polyglot Programming}
We think the move to cloud native architectures requires software engineers to rethink the criteria for how programming languages are selected.  In 2013 Meyerovich and Rabkin\cite{meyerovich2013empirical} reported on empirical human factors that impact programming language selection.  Their findings cite reasons such as open source libraries, existing code, and programmer experience as the primary drivers for selecting programming languages for new projects. To complicate matters further, in some organizations approved programming languages are standardized removing the software engineering community from the decision making loop.

The general criteria to evaluate programming languages often examines attributes like object-oriented vs functional; high-level vs low-level; type safety vs dynamic; general purpose or domain specific, and so on.  While these are good attributes to categorize programming languages they don't factor in criteria aligned to cloud native computing objectives.  For example\cite{flauzino2018you} did a comparative analysis of Java vs Kotlin.  Kotlin has been increasing in popularity within the Java community because it is less verbose, introduces modern programming language features, while interoperating well with existing Java code.  One of the criteria for good cloud native software discussed earlier is being able to move fast, thus adopting a language like Kotlin that is more productive and easier to test represents a good engineering tradeoff for organizations with significant investments and skillsets in Java. 

We think an approach for programming language selection should be based on a careful tradeoff analysis using cloud native computing architecture decisions to guide the selection.  This will often lead to a polyglot outcome, where more than one programming language is selected.  One interesting study in this space was conducted by Cordingly et. al.\cite{Serverless} where they examined the Java, Python, Go, and Node.js against a collection of different Function as a Service (FaaS) workloads.  We like their strategy using drivers such as performance and cost as the evaluation criteria.  They also used specific FaaS concerns such as cold and warm start times in their analysis. 

We think the approach used by Cordingly should be expanded to other cloud native architecture options.  For example, with container based solutions, languages like Java tend to produce very large containers, and require significant resources associated with bringing along the JVM.  Modern languages like Go, designed with the cloud in mind\footnote{Go is the primary language used to build significant cloud native platforms like Docker and Kubernetes}, produce very small containers, and have a robust and modern runtime.  Languages like Javascript and Typescript coupled with the Node.js runtime is highly optimized to support asynchronous event-based architectures. Languages like Rust provide C/C++ performance, but have a modern runtime and provide compiler-enforced memory safety. In addition to the languages themselves, additional factors such as the maturity and completeness of cloud provider supplied language-specific SDKs should also be considered in the selection criteria. 

\subsection{Multi-Region and Multi-Cloud}
The major cloud providers run massive infrastructures that have historically have provided availability that any individual enterprise would envy. However, while rare, cloud providers have had outages that have resulted in significant impact to customers. As cloud adoption continues to grow, the impact of these periodic outages will also continue to grow.  Additionally, the major cloud providers compete with each other via their innovation investments into fully managed services.  This has consequences of cloud vendor lock-in, making it hard for a customer to migrate from one cloud provider to another based on having to redevelop and redeploy their software on another providers platform.  Software engineers need to be well versed in the options and consequences from a cost and scale perspective with respect to making Multi-Region and Multi-Cloud decisions. For example: 

\begin{itemize}
	\item As mentioned in Section \ref{subsec:BasicCloudArchitecture} cloud providers offer resiliency and scale using the concept of a Region comprised of multiple Availability Zones.  Most cloud providers handle the complexity of enabling services to run across AZs in a way that is transparent to customers.  Little engineering effort is required by customers to continue to operate correctly when a single AZ fails within a region.  However, if an entire region fails, additional engineering is required to continue correct operation. These efforts increase the complexity of the software, software testing, and cost as the customer becomes responsible for data replication strategies and redundant storage costs to operate across regions. Netflix had authored engineering notes on the strategy that they use on their techblog\cite{NetflixMultiRegion}, and also created open source modules to perform chaos testing to validate their platforms ability to survive various types of failures. 
	
	\item While cloud providers offer many services that are similar, they also compete by trying to differentiate their suite of fully managed PaaS/FaaS offerings.  Software engineers can lower the blast radius of failures by diversifying services across different cloud providers.  For example, if a software engineer decided for their workloads that AWS has a superior FaaS solution with Lambda, and GCP has the best managed Kubernetes service GKE, consideration could be given to a best-of-breed strategy.  While this would reduce risk of individual cloud failure exposure, it also comes with risk and complexity.  Latency could suffer as intra-application calls have to leave one cloud and enter another.  Also, many cloud providers provide tiered pricing models, so maximizing use on one cloud provider drives the best discounts.  
	
	\item Even with being able to successfully run across different cloud providers or multiple regions software engineers must design applications that continue to operate with reduced function in the face of failure.  These tradeoffs would need careful consideration and be domain focused.  For example, on an e-commerce platform, favoring services that will allow customers to make and pay for orders can be prioritized over the ability to fulfill orders until normal cloud operations are restored. 
	
	\item Being able to run applications across cloud providers, or to port to another cloud provider requires software engineers to make careful product selection decisions.  For example, consider AWS Dynamo and Azure Cosmos.  These are both database solutions that are often compared to each other with respect to resilience, hyper-scalability and performance.  From a software interface perspective they are very different and would require a significant rewrite to port from one solution to another.  Choosing an alternative technology like managed PostgreSQL for databases would be easier to support across different cloud providers.  Kubernetes is another example of a platform that will be more similar to support across different cloud providers.  In all cases, the processes to deploy and secure cloud assets across different cloud providers is not the same, so adopting a multi-cloud strategy, even with similar technologies, will still introduce cost and complexity.
\end{itemize}  

\section{Cloud Computing Education}
One area we think is underserved by the research community is investigation into software engineering specific cloud education.  Queries of Google Scholar\cite{GoogleScolar} for research on Cloud Native Software Engineering produce very few results of relevance.  Most of the results focus on work to advance specific technical areas within the field of software engineering, for example, running AI/ML workloads, automation, software testing, or supporting microservice based architectures.  

So how do software engineers learn cloud computing?  The cloud providers themselves have done a good job filling part of the need via education and certification programs.  Cloud certifications are highly valued by industry and software engineers themselves, as they often post their certification accomplishments on personal LinkedIn pages. While cloud certifications are valuable, they focus on how to accomplish activities on a cloud platform versus how to engineer good software products using cloud services.  We have discussed FaaS as an important cloud software engineering opportunity earlier. Certifications would enable engineers to understand how to deploy a FaaS component on AWS Lambda, Azure Functions, or Google Cloud Functions, but they would not cover important other considerations such is your architecture event-driven, how will state be managed, SDK robustness, component granularity ({\em e.g.}, is a function too small) and so on.

We think an important research question is to address if academic institutions play a leadership or partnership role to address the educational needs of software engineers working in the cloud.  It appears that most of the collective knowledge in this space is acquired with on-the-job experience and via publications authored by industry professionals.  

\section{Conclusion}
This paper brings attention to the need for additional focus on expanding software engineering practices given the trend of moving to cloud and edge computing.  We examined many cloud computing architectural concepts from the lens of a software engineering practitioner and summarized many opportunities that would benefit the community. Many organizations transition to the cloud by taking their top engineers and focusing them on moving existing digital assets like web and mobile applications to the cloud.

Early successes, coupled with the attractive technical capabilities liked by engineers, and a pay-as-you-go model liked by managers continue to drive acceleration of the cloud.  We think this next wave of cloud will require more software engineering rigor as we need to scale the number of qualified engineers to work in the cloud, while at the same time, needing to support moving core enterprise applications to the cloud that don't have the same architectural characteristics as digital applications. 

We also discussed new business opportunities that could only emerge with the cloud such as API-based products, and edge computing.  These are complex architectures that will require software engineers to master additional skills.

%\section{Acknowledgements}
%We would like to thank the following individuals for their valuable feedback %to this paper. Your name can go here. 

\bibliography{CNSE-Arxiv-Preprint-Mitchell.bib}

\iftrue
\vspace{1cm}

\begin{wrapfigure}{l}{0.35\columnwidth}
\vspace{-\intextsep}
\includegraphics[width=1.0\linewidth]{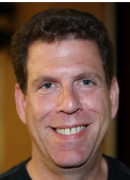}
\vspace{-20pt} 
\label{fig:BrianMitchell}
\end{wrapfigure}

\noindent\textbf{Brian S. Mitchell} is an accomplished technologist, engineer, educator, software engineering researcher, speaker, strategist, leader, and enterprise-scale change agent. Brian is currently a member of the Department of Computer Science at Drexel University. His career has spanned both industry and academia, including holding the Distinguished Engineer role at a Fortune 15 company. He provided technical thought leadership and directed teams responsible for driving disruptive digital innovation that led to the creation of multiple generations of products that help millions of people every day. Brian also has more than 20 years of teaching experience in a variety of areas including Software Engineering, Software Architecture, Operating Systems, Networks, Computer Architecture, Programming Languages, and Distributed Systems. His recent research interests include exploring several interesting problems at the intersection of Software Engineering, Software Architecture and Cloud Native Computing. Previously he was one of the founders of the Search-Based Software Engineering research space, publishing many influential papers focused on recovering software architecture insights directly from source code. Dr. Mitchell holds BS, MS and PhD degrees in Computer Science, and a ME in Computer \& Telecommunication Engineering.

\fi
\end{document}